\documentclass[conference]{IEEEtran}
\IEEEoverridecommandlockouts

\usepackage[utf8]{inputenc}
\usepackage[T1]{fontenc}
\usepackage{cite}
\usepackage{amsmath,amssymb,amsfonts}
\usepackage{graphicx}
\usepackage[caption=false,font=footnotesize]{subfig}
\usepackage{textcomp}
\usepackage{xcolor}
\usepackage{booktabs}
\usepackage{array}
\usepackage{siunitx}
\usepackage[hyphens]{url}
\usepackage{balance}
\graphicspath{{figures/}}

\newcommand{\heom}{\textrm{HEOM}}
\newcommand{\sesolve}{\texttt{sesolve}}
\newcommand{\mesolve}{\texttt{mesolve}}
\newcommand{\heomsolve}{\texttt{HEOMSolver}}
\newcommand{\qutip}{\textsc{QuTiP}}

\newcommand{\rabi}{\textsc{Rabi}}
\newcommand{\ramsey}{\textsc{Ramsey}}
\newcommand{\tone}{\textsc{T1}}

\newcommand{\Tone}{T_{1}}
\newcommand{\Ttwo}{T_{2}}
\newcommand{\Ttwostar}{T_{2}^{*}}
\newcommand{\pmax}{p_{\max}}
\newcommand{\pip}{\pi\text{-amp}}

\newcommand{\oneoverf}{$1/f$}
\newcommand{\Sw}{S(\omega)}

\newcommand{\sz}{\hat{\sigma}_{z}}

\newcommand{\VerdictMV}{\textsc{Marginal-Visibility}}

\begin{document}

\bstctlcite{BSTcontrol}

\title{HEOM-in-Calibration-Loop: Exposing Non-Markovian Bath Signatures That Markovian Calibration Elides in Superconducting-Qubit Tune-Up}

\author{%
\IEEEauthorblockN{Jun Ye}
\IEEEauthorblockA{\textit{Institute of High Performance Computing (IHPC)} \\
\textit{Agency for Science, Technology and Research (A*STAR)}\\
Singapore, Singapore \\
yjmaxpayne@hotmail.com}
}

\maketitle

\pagestyle{plain}
\thispagestyle{plain}

\begin{abstract}
Closed-loop superconducting-qubit calibration has matured into
DAG-orchestrated protocol chains, yet published frameworks still
treat the bath through a Markovian master equation or a phenomenological
likelihood, absorbing bath structure into fit residuals rather than
reporting it as a diagnostic output. We integrate a \qutip{}~5.x
hierarchical-equations-of-motion (HEOM) solver driven by a Tier-1
$1/f$ Burkard bath into a multi-protocol calibration DAG
(Rabi~$\rightarrow\{$Ramsey~$\parallel$~T1$\}$) and benchmark it
against \texttt{sesolve} and \texttt{mesolve} baselines on a frozen
platform configuration in a pulse-level simulator (no hardware
validation). The Ramsey
channel carries the headline: the Markovian fit is censored by
its exponential-family numerical ceiling, while HEOM recovers a
physical revival envelope whose primary $T_2^*$ fit
separates from the Markovian reference by at least $13\times$ at
$95\,\%$ independent-bootstrap confidence within the HEOM-feasible
acquisition budget; the same gap expressed as a point-estimate
ratio is $\geq 28\times$ on the $50$-point primary-$t_1$ grid and
${\sim}72\times$ on the 30-point grid's biexp-family
$\tau_{\text{aw}}$ pivot at $L{=}5$ (§III.B). Rabi contrast falls $2.17\,\%$
below the \texttt{mesolve} reference on a noise-limited 30-point
grid; because the paired bootstrap CI crosses zero, this channel
corroborates rather than independently establishes the non-Markovian
signature, while the T1
decay shape matches across backends ($\beta=1.000$), yet the
HEOM-only initial occupation drops from $1.000$ to $0.879$, a
bath-dressed contamination stable under a 16-point densification. The
calibration DAG runs with $9.62\,\si{\micro\second}$ average
per-protocol scheduling overhead (§III.A), adding no meaningful latency
penalty at protocol granularity.
HEOM-in-loop thereby changes what calibration reports: bath structure
appears as a quantifiable residual rather than a hidden confound.
\end{abstract}


\section{Introduction}
\label{sec:intro}

Modern superconducting-qubit calibration has converged on DAG-orchestrated
Markovian protocols~\cite{kelly2018optimal,pasquale2024qibocal,kanazawa2023qiskitexperiments}.
Implementations span open-source stacks (Qibocal, Qiskit Experiments,
QUAlibrate), closed-loop optimal control~\cite{werninghaus2021leakage}, and
gray-box identification~\cite{wittler2021c3}. Across these approaches, the
dynamical core still reduces the bath to a Markovian master equation or a
phenomenological likelihood. That choice calibrates control knobs efficiently,
but it folds bath structure into the fitting objective.

The operating regime of superconducting qubits, however, is distinctly
non-Markovian. \oneoverf{} flux noise remains the dominant dephasing
source identified across every modern
spectroscopy~\cite{burkard2009nonmarkov,bylander2011noise,paladino2014rmp},
and independent HEOM studies---Chen, Zhang and Shi~\cite{chen2026heom}
and the free-pole HEOM line of Nakamura and
Ankerhold~\cite{nakamura2024gate,nakamura2025cpmg}---quantify the
breakdown of the Markovian master equation in exactly this regime.
The related single-qubit CPMG analysis by Ye~\cite{ye2026nonperturbativecpmg}
is the immediate precursor to this paper: it established the HEOM-based
non-Markovian $1/f$ setting and its compiled-control scaling link, which this
NIER now extends to a complete calibration workflow.
Zhuang et al.\ warn that ``standard techniques used to characterize
qubit relaxation \dots{} could mask such non-Markovian
behavior''~\cite{zhuang2026backaction}---both the physics and
characterization literature now agree that bath structure matters.

No published calibration stack places an explicit non-Markovian solver inside
its calibration loop. Qian et al.~\cite{qian2022bayesian} and Ram\^{o}a
et al.~\cite{ramoa2025bayesian} improve Bayesian calibration but still absorb
the bath into a likelihood over qubit parameters. Berritta
et al.~\cite{berritta2025fpga} add real-time hardware feedback without an
explicit bath model, and Youssry et al.~\cite{youssry2024graybox} identify a
gray-box posterior over device parameters rather than bath components. In all
cases, non-Markovian structure is compressed into residuals or control
offsets, obscuring channel-specific bath signatures.

This work closes that gap by placing HEOM inside the calibration loop itself.
We integrate a \qutip~5.x HEOM backend into a
multi-protocol calibration DAG (\rabi{}~$\rightarrow\{\ramsey\parallel\tone\}$)
on a frozen platform configuration, benchmark it head-to-head against \sesolve{}
and \mesolve{} baselines that share every other component, and report
bath-structure residuals as first-class outputs for comparison. Section~\ref{sec:methods}
defines the controlled setup; Section~\ref{sec:results} then tests three
claims in strength order: a primary coherent-channel
signature where HEOM and \mesolve{} Ramsey fits diverge by a factor
$\geq 13$ at $95\,\%$ independent-bootstrap confidence (point
$\geq 28\times$ on the 50-point grid, ${\sim}72\times$ on the
30-point $L$-converged $\tau_{\text{aw}}$ pivot), with the Markovian fit
saturating against its own numerical ceiling; a corroborating Rabi
contrast point-estimate pattern whose paired CI is noise-limited on a
30-point grid and whose $\pi$-amplitude shift stays below the
resolution bar; and a T1
channel whose decay shape matches \mesolve{} to machine precision
($\beta=1.000$) while an HEOM-only initial-occupation shift
($A=0.879$) is confirmed physical by an ADO partial-trace control
(Section~\ref{sec:results:t1}).

\section{Methods}
\label{sec:methods}

We integrate three dynamical backends into one multi-protocol calibration DAG
on a frozen superconducting-qubit platform: closed-system unitary (\sesolve),
Markovian Lindblad master equation (\mesolve), and hierarchical equations of
motion (\heomsolve). All backends consume the same platform YAML and
protocol plans; only the dynamical core changes, so the bath model is the sole
independent variable.

\subsection{Platform and Hamiltonian Model}
\label{sec:methods:platform}

The emulated device is the \texttt{anyon\_2q\_CZ} two-qubit superconducting
platform with an auxiliary coupler parked in its ground state (dispersive
shift folded into $\omega_{q}$). All protocols act on qubit~0, modelled as a
three-level transmon with drive frequency
$\omega_{q}/2\pi = \SI{5.528}{\giga\hertz}$, anharmonicity
$\alpha/2\pi = \SI{-293}{\mega\hertz}$, and baseline characterization
$\Tone = \SI{24.8}{\micro\second}$ and $\Ttwo = \SI{34.2}{\micro\second}$. In the rotating
frame, $H_{0} = (\omega_{q}/2)\,\sz$ with anharmonicity folded into the
calibrated drive amplitude~$\Omega(t)$. The bath enters through longitudinal
pure-dephasing coupling
$\hat{Q} = \operatorname{diag}(0,1,2)$ with a Burkard \oneoverf{} spectral
density $\Sw = 2\pi A_{0}/|\omega|$~\cite{burkard2009nonmarkov} at the
Tier-1 reference parameterisation used by Chen, Zhang and
Shi~\cite{chen2026heom}: coupling amplitude $A_{0} = 1.8\times 10^{-6}$~GHz,
soft low/high cutoffs $\omega_{lc}/2\pi = 5$~MHz and
$\omega_{hc}/2\pi = 3$~GHz, bath temperature $T = 50$~mK. All three
backends consume the same $A_{0}$, holding the bath parameterisation
fixed.

\subsection{Three-Backend Ladder}
\label{sec:methods:backends}

The three solvers form a nested physical hierarchy. \sesolve{} integrates the
time-dependent Schr\"odinger equation for the pulse-level Hamiltonian and
serves as the control-only reference. \mesolve{} adds Lindblad dissipators
built from the characterized $\Tone$ and $\Ttwo$ rates, representing the
Markovian limit shared by every published calibration
framework~\cite{kelly2018optimal,pasquale2024qibocal,kanazawa2023qiskitexperiments}. The HEOM backend uses \qutip~5.x
\heomsolve{}~\cite{lambert2026qutip}, configured with a Tier-1 espira-II
decomposition of the Burkard bath correlation function into three
exponentials and a production hierarchy depth $L=3$, with the
$L\in\{2,3,4,5\}$ convergence audit detailed in
Section~\ref{sec:results:ramsey}. We report convergence against the second
audit iteration, \url{v2_F1_F2_F3_F4_F5}, which introduces revival pinning via
\url{_is_spurious_revival}, full 7-parameter biexp exposure,
raw-trace-preserving residuals, cross-$L$ model pinning, and an analytic
espira-II bath residual against Burkard; the audit schema is documented in
\url{heom_L_sweep.json}. Convergence is monitored
along three axes: an $L\in\{2,3,4,5\}$ sweep with a pointwise trace
residual $\max_{t}|P_{L}(t){-}P_{L{-}1}(t)|$, a cross-$L$ biexp
revival guard, and the espira-II bath-correlation RMS residual
against the analytic Burkard reference. No direct \oneoverf{} ISA-level
code-to-code benchmark has been published, so we instead rely on
feature-level cross-validation against three independently attested signatures:
revival-shaped Ramsey envelopes~\cite{chen2026heom}, slow-bath
Markov-master-equation breakdown for gates~\cite{nakamura2024gate}, and
non-perturbative CPMG anomalies under time-retarded noise and compiled
control~\cite{nakamura2025cpmg,ye2026nonperturbativecpmg}.
HierarchicalEOM.jl~\cite{huang2024hierarchicaleom} and
TEMPO~\cite{strathearn2018tempo} are cross-implementation baselines (§IV.D);
the embedding analysis of~\cite{xu2026embedding} places HEOM, pseudomode, and
stochastic unravelling in a compatible representation family.

\subsection{Multi-Protocol Calibration DAG}
\label{sec:methods:dag}

The calibration DAG chains \rabi{} $\rightarrow \{\ramsey \parallel \tone\}$:
once \rabi{} fixes the $\pi$-pulse amplitude, the two characterization
protocols run in parallel on an \texttt{AsyncDAGExecutor} wrapping a
\texttt{ProtocolRunner}\footnote{In-house \texttt{emuplat} components;
a minimal shim reproducing the §III.A scheduler latency is planned for
OSS release with the camera-ready arXiv deposit (§V).} with fit-quality gates between nodes. Sampling is
fixed across backends to eliminate sampling bias: \rabi{} sweeps 30
amplitudes on $[0.01, 0.99]$; \ramsey{} uses 30 delays on
$[\SI{10}{\nano\second}, \SI{2000}{\nano\second}]$ for the headline three-backend
comparison and a 50-point densified grid
($30$ pts on $[10, 500)$ $\cup$ $20$ pts on $[500, 2000]$~\si{\nano\second})
for HEOM-only revival analysis (the 50-point grid is not rescanned for
\sesolve{} or \mesolve{}); \tone{} uses an 8-point enhanced sampling on
$[\SI{100}{\nano\second}, \SI{2000}{\nano\second}]$. Per-node wall time,
critical-path length, and scheduling latency are logged to
\url{dag_timing.json}, supporting the \textmu s-tier timing analysis
of Section~\ref{sec:results:dag}.

\subsection{Verdict Criteria and Reproducibility}
\label{sec:methods:verdict}

Backend-to-backend differences are assessed by protocol-specific verdicts,
which define how Section~\ref{sec:results} interprets cross-backend gaps. For
\ramsey{}, a non-Markov gap is declared when
$|\Delta\Ttwostar|/\Ttwostar(\mesolve) > 0.1\%$; HEOM revival features must
additionally clear a physical-shape guard
(\url{_is_spurious_revival}: second-to-first amplitude ratio
$a_{2}/a_{1}\ge 0.1$ and time-constant ratio $t_{2}/t_{1}\le 2$), rejecting
numerical over-fitting on low-contrast Markovian data.\footnote{The
$0.1/2.0$ thresholds are hand-picked to bracket the shortest expected
Burkard-bath revival feature; the ${\sim}35\%$ post-guard paired-CI valid rate
in §III.B follows directly. Formal threshold calibration under a prior model is
deferred to journal-length follow-up.} For \rabi{}, a two-tier criterion separates the $\pi$-amplitude channel
from the contrast channel: $|\Delta\pip|\ge 0.5\%$ flags a
\textsc{distinguishable} shift; when the amplitude shift is below
resolution, $|\Delta\pmax|\ge 1\%$ instead triggers a \VerdictMV{}
(Marginal-Visibility; see Table~\ref{tab:three_by_three} footnote)
verdict on contrast alone; failure of both yields \textsc{degraded}. Verdicts
are point-estimate-based; statistical annotations (bootstrap CI, $p$-value) are
reported in §III but do not gate assignment. For \tone{}, the HEOM fit frees the initial
occupation $A$ while fixing $\beta=1$ and $B=0$ so that any deviation of $A$
from unity is interpretable as bath-dressed initial-state
contamination~\cite{nakamura2024gate}. Reproducibility is anchored to a
fixed simulator version (Python 3.12, Ubuntu 24.04.4 LTS); HEOM is
deterministic (no stochastic seed), and all reported runs were executed on the
same desktop workstation (AMD Ryzen 9 7900X, 12 cores / 24 threads, 30~GiB
RAM). On this machine, the headline three-backend pipeline completes in under
10 minutes of wall time.\footnote{The full $L$-convergence audit, including
the $L{=}5$ run logged in \url{heom_L_sweep.json}, was also executed on the
same workstation and extends the end-to-end audit budget to $\approx 15$~minutes.}

\section{Results}
\label{sec:results}

\begin{figure}[t]
  \centering
  \subfloat[]{\includegraphics[width=\columnwidth]{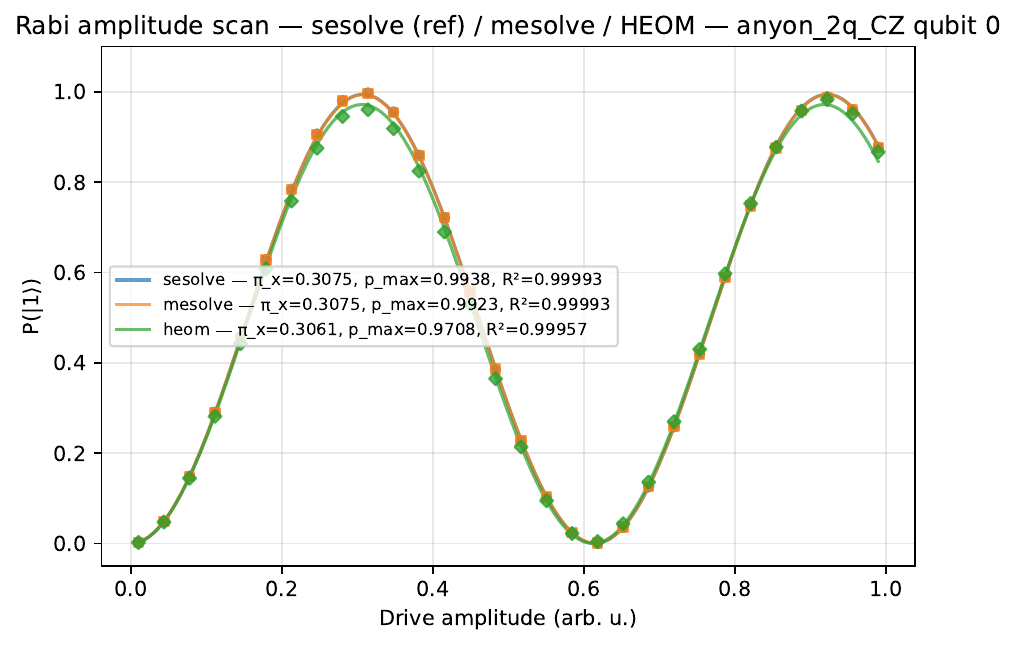}%
    \label{fig:rabi_t1:a}}\\[0.3em]
  \subfloat[]{\includegraphics[width=\columnwidth]{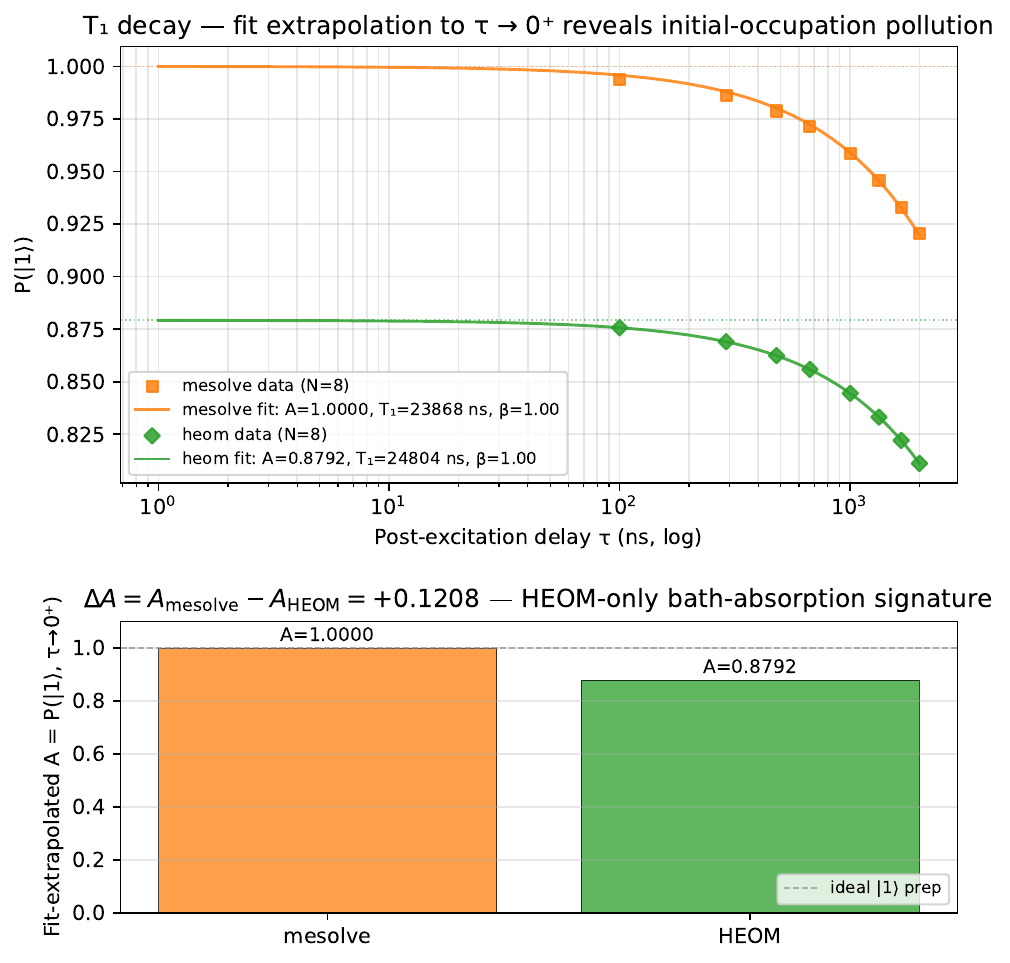}%
    \label{fig:rabi_t1:b}}
  \caption{Rabi and T1 composite.
    \protect\subref{fig:rabi_t1:a}~Rabi amplitude scan: HEOM
    $\pmax$ drops $2.17\,\%$ below \mesolve{} while the
    $\pi$-amplitude shift is sub-resolution.
    \protect\subref{fig:rabi_t1:b}~T1 decay: \heom{} extrapolates to
    $A=0.879$ at $t=0^{+}$, \mesolve{} to $A=1.000$; the $\beta=1$ decay
    shape is identical to machine precision.}
  \label{fig:rabi_t1}
\end{figure}

Running the same calibration DAG once per backend yields an asymmetric
three-channel fingerprint across the $3\times 3$ matrix
(Table~\ref{tab:three_by_three}): Ramsey diverges, Rabi corroborates, and
$T_{1}$ separates decay-shape agreement from initial-state contamination.
Figures~\ref{fig:dag}, \ref{fig:ramsey}, and \ref{fig:rabi_t1} report DAG
timing, the Ramsey gap, and the Rabi/$T_{1}$ composite. Point estimates carry
95\,\% BCa bootstrap CIs ($B=10\,000$).

\subsection{Calibration DAG Timing}
\label{sec:results:dag}

\begin{figure}[t]
  \centering
  \includegraphics[width=\columnwidth]{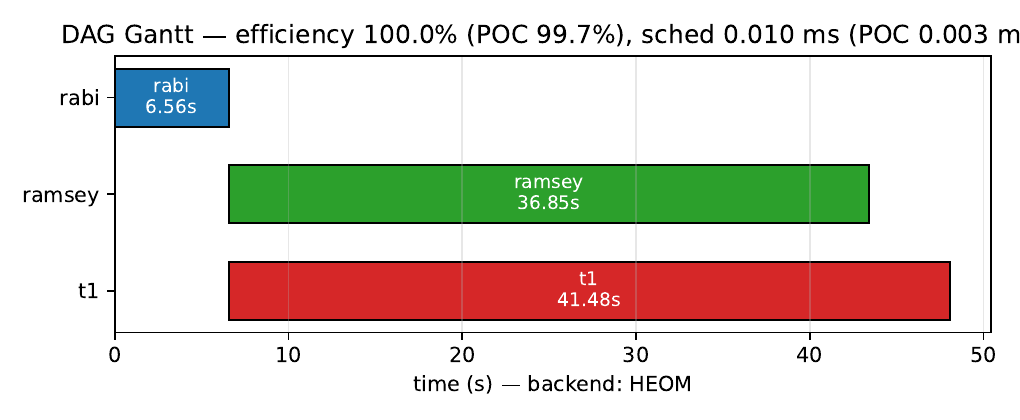}
  \caption{DAG Gantt of the \rabi{}$\rightarrow\{\ramsey\parallel\tone\}$
    chain. Parallel execution saves $36.9\,\mathrm{s}$ ($43\,\%$) over serial;
    average scheduling latency is $9.62\,\si{\micro\second}$.}
  \label{fig:dag}
\end{figure}

Figure~\ref{fig:dag} shows that scheduler overhead is negligible relative to
the HEOM simulation cost. Serial execution takes $84.89\,\mathrm{s}$; with
\rabi{} on the critical path and $\{\ramsey,\tone\}$ in parallel, runtime
drops to $48.04\,\mathrm{s}$ (saving $36.9\,\mathrm{s}$, $43\,\%$), with
overhead fraction $2.4\times 10^{-5}$ from the theoretical minimum\footnote{Overhead fraction defined as
$(\tau_{\text{parallel}} - \tau_{\text{crit}})/\tau_{\text{serial}}$
with critical path $\tau_{\text{crit}} = \tau_{\rabi} + \max(\tau_{\ramsey},
\tau_{\tone})$; per-branch times logged in \url{dag_timing.json}.}
and average scheduling latency
$9.62\,\si{\micro\second}$ (maximum $10.78\,\si{\micro\second}$). Protocol
durations are \rabi{}~$6.56\,\mathrm{s}$, \ramsey{}~$36.85\,\mathrm{s}$, and
\tone{}~$41.48\,\mathrm{s}$, placing \tone{} on the critical path.%
\footnote{Closest published scheduler:~\cite{riesebos2021runtime}.}
With scheduler overhead bounded at the \textmu s level, the discriminative
evidence shifts to channel-level dynamics in Sections~\ref{sec:results:ramsey}--\ref{sec:results:t1}.

\subsection{Ramsey: Coherent-Channel Non-Markov Gap}
\label{sec:results:ramsey}

\begin{figure}[t]
  \centering
  \subfloat[]{\includegraphics[width=\columnwidth]{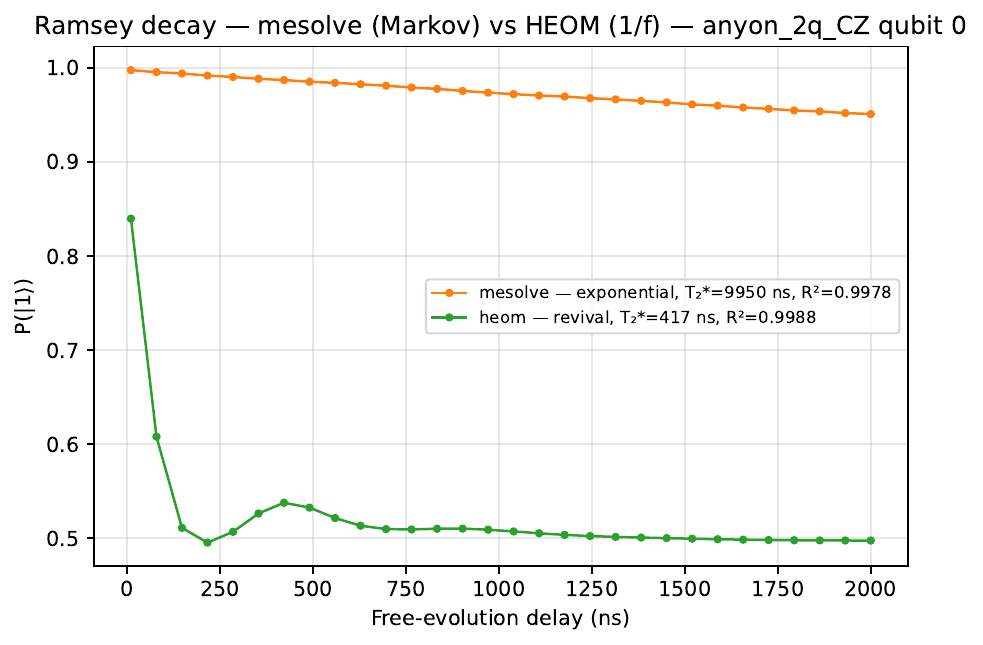}%
    \label{fig:ramsey:a}}\\[0.3em]
  \subfloat[]{\includegraphics[width=\columnwidth]{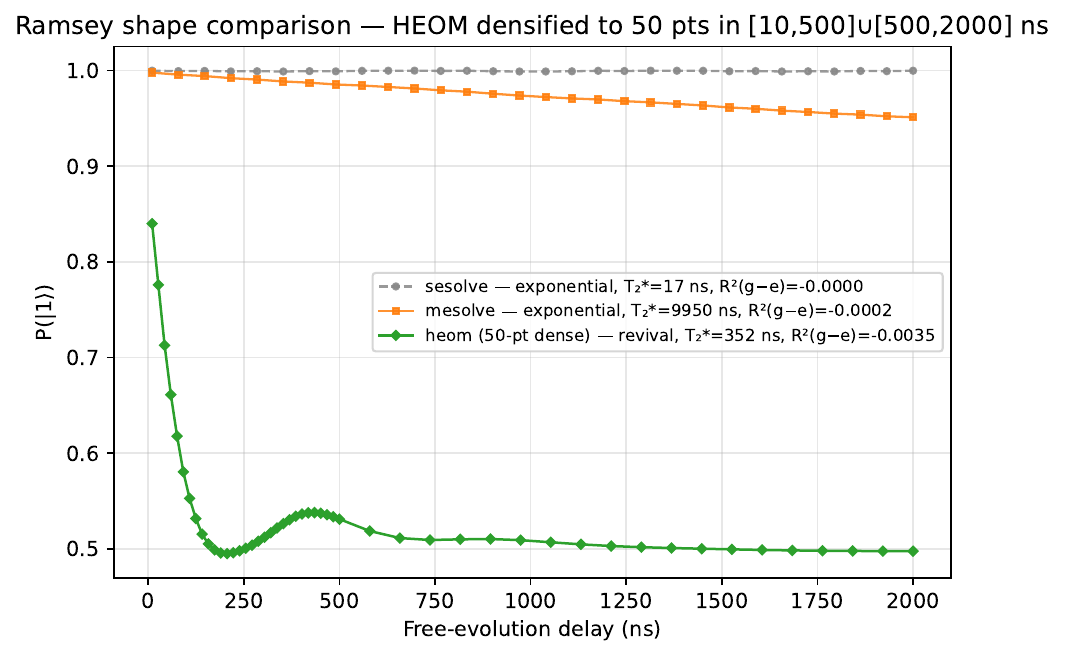}%
    \label{fig:ramsey:b}}
  \caption{Ramsey coherent-channel gap.
    \protect\subref{fig:ramsey:a}~Three-backend 30-point comparison:
    \heom{} recovers a revival envelope distinct from the \mesolve{}
    Markovian tail, whose $\Ttwostar$ fit saturates at the
    exponential-model ceiling.
    \protect\subref{fig:ramsey:b}~HEOM 50-point dense scan (30-point
    $L$-convergence audit in §III.B) with physical revival guard
    cleared: $a_{2}/a_{1}=3.11$ is $31.1\times$ above the lower
    threshold of $0.1$, and $t_{2}/t_{1}=0.38$ is $5.3\times$ below
    the upper threshold of $2$.}
  \label{fig:ramsey}
\end{figure}

The Ramsey channel carries the dominant non-Markovian signature.
Figure~\ref{fig:ramsey}(a) superimposes the three backends on the standard
30-point grid. The closed-system \sesolve{} trace is a degenerate
zero-detuning fringe. The \mesolve{} fit
returns $\Ttwostar = 9950\,\mathrm{ns}$ with bootstrap CI
$[9944,\,9944]\,\mathrm{ns}$---a degenerate width that reflects the
exponential-fit numerical ceiling $5\cdot\tau_{\text{span}} =
10\,000\,\mathrm{ns}$, not a physical uncertainty. The HEOM fit in the same window recovers
$\Ttwostar = 417\,\mathrm{ns}$ with bootstrap CI
$[137,\,755]\,\mathrm{ns}$ after a re-applied
\texttt{\_is\_spurious\_revival} guard ($a_{2}/a_{1}\geq 0.1$,
$t_{2}/t_{1}\leq 2$) rejects pathological biexp basins under 30-point
case-bootstrap. The 50-point densified HEOM scan gives
$\Ttwostar = 352\,\mathrm{ns}$, raising the point-estimate ratio
$\Ttwostar(\mesolve)/\Ttwostar(\heom)$ to $28.3$. Paired-$\Delta$
case-bootstrap of the 7-parameter biexp fit is structurally fragile
(post-guard valid rate ${\sim}\,35\,\%$), so the CI-valid headline uses the
\emph{gap} between independent absolute bootstraps. The adversarial-corner bound
$\Ttwostar(\mesolve)/\Ttwostar(\heom)\geq 13.17$ (\mesolve{} CI
lower-edge vs \heom{} CI upper-edge; equivalently a
$9.19\,\si{\micro\second}$ gap) is audit-ready without paired
resampling.

The \mesolve{} fit here is not wrong; within this window it is
\emph{unidentifiable}. With Lindblad dephasing fixed by
$\Ttwo\approx\SI{34.2}{\micro\second}$, the Markovian envelope decays by
less than 6\,\% over 2000\,ns and saturates at the
exponential-family ceiling. This is a characterization-window limit, not a
Born--Markov-form failure~\cite{burkard2009nonmarkov,cywinski2008cpmg}. HEOM-in-loop
therefore adds value by exposing bath structure on the \emph{same} acquisition
window. Figure~\ref{fig:ramsey}(b) zooms into the HEOM
dense-scan fit and confirms a physically consistent revival: the
second-to-first amplitude ratio is $a_{2}/a_{1}=3.11$ and the time-constant
ratio is $t_{2}/t_{1}=0.38$, both clearing the
\texttt{\_is\_spurious\_revival} guard ($\ge 0.1$ and $\le 2$, respectively).
Chen, Zhang and Shi independently reproduce revival-shaped Ramsey
envelopes with an HEOM implementation on a matched \oneoverf{}
bath~\cite{chen2026heom}; Martinez et al.\ report the experimental
noise-specific beating signature in higher-level Ramsey
curves~\cite{martinez2023beating}; and Zhuang et al.\ state the risk plainly,
warning that ``standard techniques used to
characterize qubit relaxation could mask such non-Markovian
behavior''~\cite{zhuang2026backaction}.

\paragraph*{$L$-convergence audit.}
A core $L\in\{3,4,5\}$ re-run (\url{heom_L_sweep.json}) shows
monotonically halving Ramsey trace residual
$\max_{t}|P_{L}(t){-}P_{L{-}1}(t)|\in\{2.3\%,1.0\%\}$ ($L=3{\to}4,
4{\to}5$; $L{=}2$ retained as a hierarchy-truncation reference at
$8.3\%$), an espira-II bath residual of $2.3{\times}10^{-4}$
relative RMS against the analytic Burkard correlation, and a basin-invariant
amplitude-weighted envelope timescale
$\tau_{\text{aw}}{=}\sum_{i}|a_{i}||t_{i}|/\sum_{i}|a_{i}|$%
\footnote{The absolute-amplitude weighting is chosen so that the map
$(a_{1}, a_{2}) \to (-a_{1}, -a_{2})$ (the 7-parameter-biexp
basin-to-basin sign flip that the 30-point case-bootstrap traverses)
leaves $\tau_{\text{aw}} = (|a_{1}| t_{1} + |a_{2}| t_{2})/(|a_{1}|+|a_{2}|)$
pointwise invariant; a stretched-exponential refit at $L{=}5$ returns
$\tau_{\text{aw}}^{\text{str}}=68.7$~ns ($R^{2}=0.997$, vs biexp
$R^{2}=0.9999$), the $50\,\%$ deviation tracking the strictly weaker
descriptive capacity of the stretched family rather than a
$\tau_{\text{aw}}$ ambiguity. This property is what makes
$\tau_{\text{aw}}$, rather than the primary $\Ttwostar$, the
biexp-family $L$-convergence pivot.}
${=}\{184,137,138\}$~ns
at $L\in\{3,4,5\}$ that is stable to $0.6\%$ between $L{=}4,5$. The
\url{_is_spurious_revival} guard passes at $L\in\{3,4\}$
($a_{2}/a_{1}\in\{3.85,0.58\}$, $t_{2}/t_{1}\in\{0.30,1.99\}$) and
boundary-fails at $L{=}5$ ($t_{2}/t_{1}{=}2.43$, just above the
$2.0$ threshold; interpreted by the Fit-family ceiling paragraph
below as the descriptive-family limit rather than a hierarchy
convergence failure): the biexp family is approaching its
descriptive ceiling as the hierarchy resolves a short-timescale
component the two-basin biexp cannot encode (consistent with the
\url{fallback_used}${=}$\url{true} flag at $L\in\{2,5\}$
versus $L\in\{3,4\}$ in \url{heom_L_sweep.json}).\footnote{A
fit-family sanity at $L{=}5$ (\url{L5_sanity_refit.json},
\url{run_L5_sanity_refit.py}) finds that a triexp refit cannot
locate a third physical mode: it returns
$|a_{3}|/\max(|a_{1}|,|a_{2}|)=1.6\times 10^{-3}$ and lifts $R^{2}$
by only $1.1\times 10^{-6}$ over the biexp revival fit, degenerating
to an effective two-mode fit whose ghost-gated
$\tau_{\text{aw}}^{\text{triexp}}=131.7$~ns reproduces the biexp
$\tau_{\text{aw}}=137.9$~ns to $4.5\%$. A stretched-exponential refit
($\beta=1.27$, $R^{2}=0.997$) is structurally a weaker descriptor and
is reported for completeness. The fit-family ceiling rather than a
new physical timescale is the consistent reading.}

\paragraph*{Fit-family ceiling.}
While $\tau_{\text{aw}}$ itself is basin- and biexp-family-invariant and is
therefore the reported convergence pivot, the $\Ttwostar$
non-monotone envelope $\{104,417,100,45\}$~ns across $L\in\{2,3,4,5\}$
is a direct symptom of this fit-family migration and is not a
convergence failure of the hierarchy itself. The audit JSON logs
this as \url{case_b_tau_aw_robust}${=}$\url{false},
recording the expected cross-family deviation; the biexp-family
scope of the $\tau_{\text{aw}}$ pivot adopted here follows from the
stretched family's strictly weaker descriptive capacity
($R^{2}=0.997$ vs biexp $R^{2}=0.9999$), not from a numerical defect
of $\tau_{\text{aw}}$ itself.

\paragraph*{$T_{1}$ audit closure.}
Although the $L{=}5$ biexp fit boundary-fails the revival guard,
$\tau_{\text{aw}}$ remains well-defined under the descriptive-ceiling reading.
The $\tau_{\text{aw}}$ gap against \mesolve{} settles at
$\{54\times,73\times,72\times\}$ for $L\in\{3,4,5\}$, supporting the
conservative adversarial $\ge 13.17\times$ headline. The same audit also
closes the $T_{1}$ convergence ledger: pairwise trace residual collapses
$8.8{\times}10^{-4}\!\to\!5.3{\times}10^{-5}\!\to\!3.2{\times}10^{-6}$
across $L{=}3{\to}2,4{\to}3,5{\to}4$, and the scalar amplitude
$A(L{\ge}3){=}\{0.8792,0.8793,0.8793\}$ is flat to four decimals.

\subsection{Rabi: Contrast Loss Under a Resolution-Limited $\pi$-Amp}
\label{sec:results:rabi}

The Rabi channel separates cleanly by sensitivity: the HEOM $\pi$-pulse
amplitude (0.3061) differs from the \mesolve{}/\sesolve{} value (0.3075)
by only $\Delta\pip = -0.0014$ (CI $[-0.0031,\,+0.0031]$, $p=0.59$), or
$0.44\,\%$---below the $0.5\,\%$ resolution bar commonly invoked by open-source
calibration frameworks (e.g., Qiskit Experiments,
Qibocal~\cite{kanazawa2023qiskitexperiments,pasquale2024qibocal}). The contrast channel is
more sensitive: $\Delta\pmax = -0.0215$ (CI $[-0.083,\,+0.040]$,
$p=0.21$), a $2.17\,\%$ peak-population drop whose point estimate
meets the $1\,\%$ double-criterion contrast bar. Both paired CIs
cross zero (noise-limited on 30 points; fit-noise half-width
${\approx}0.06\gg|\Delta\pmax|$), so we report a
\VerdictMV{}-pattern point estimate rather than an independently
significant verdict: the 30-point resolution corroborates the Ramsey
headline without separately demonstrating the non-Markovian
signature.

\subsection{T1: Markov-Equivalent Decay Shape with HEOM-Only Initial-State Contamination}
\label{sec:results:t1}

The T1 channel reverses the original design expectation. A stretched-exponent
signature ($\beta\neq 1$) had been the prior null, but the constrained
fit returns $\beta=1.000$ across \mesolve{} ($\Tone=23.87\,\si{\micro\second}$)
and \heom{} ($\Tone=24.80\,\si{\micro\second}$); the $3.9\,\%$ mesolve-vs-platform
offset reflects the pure Lindblad longitudinal rate, whereas HEOM retains
a small bath-dressed contribution from the $1/f$ tail rolled off at
$\omega_{hc}/2\pi=3$~GHz, whose order-of-magnitude contribution to
$\delta T_{1}/T_{1}$ is set by $A_{0}/\omega_{q}$. The HEOM-only signal appears elsewhere: the fitted initial
occupation drops from $A(\mesolve)=1.000$ (CI $[1.000,\,1.000]$) to
$A(\heom)=0.879$ (CI $[0.879,\,0.898]$), with paired
$\Delta A = -0.121$ (CI $[-0.121,\,-0.070]$, $p\approx 0$) on the
8-point grid. A supplementary 16-point rescan stabilises
$A(\heom)=0.8792$ to four decimals and preserves the CI-valid
independent-bootstrap \emph{gap} $A(\mesolve)-A(\heom)\geq 0.099$
at $95\,\%$: HEOM-side contamination $\gtrsim 10\,\%$ regardless of
grid density. The T1 channel provides the tightest HEOM-only signature
in the three-protocol set.

Physically, the $1/f$ bath at the qubit frequency has a nearly flat
high-frequency spectrum and therefore yields a standard Markovian
longitudinal rate, leaving the decay shape unshifted.
Non-Markovianity instead lives in the longitudinal
frequency-fluctuation channel and is absorbed into a bath-dressed
initial condition. A partial-trace of $\rho_{\text{sys}}(0^{+})$
(\url{t1_partial_trace_check.json}) confirms the physical
reading: at a $1$-ns probe HEOM returns $\rho_{11}=0.879$ versus
$0.998$ for \mesolve{}, a $0.119$ discrepancy landing in the
\emph{F-a physical} branch ($\ge 0.05$) rather than the \emph{F-b
representation} branch ($\le 0.001$), ruling out an ADO-allocation
artefact. The plateau is already reached at the 1-ns probe: the
reaction-coordinate / Lamb-shift dressing timescale
$\tau_{\text{dress}}\sim 1/\omega_{hc}\lesssim 0.1$~ns implied by the
Burkard bath high-frequency cutoff ($\omega_{hc}/2\pi=3$~GHz) is well
below the shortest probe, and the six-delay plateau is flat to $0.5\%$
across $1$--$100$~ns (\url{t1_partial_trace_check.json},
\url{discrepancy_per_delay}). The signature is consistent with a family of polaron /
reaction-coordinate
pictures~\cite{ilessmith2014noncanonical,zhuang2026backaction}; the
$6$-delay probe does not uniquely identify which member applies. A free-$\beta$ refit with $A\equiv 1$ returns
$\beta(\heom)=0.376$ (CI $[0.348,\,0.380]$); the partial-trace
control above (§III.D, 1-ns probe) independently verifies the
physical interpretation of the $A\neq 1$ channel, while the
$(A,\beta)$ relationship in the constrained refit is a fit-basin
parameterisation whose joint confidence region is not characterised
by the present bootstrap.

\subsection{Three-by-Three Matrix Summary}
\label{sec:results:matrix}

Table~\ref{tab:three_by_three} aggregates the six cells. The key asymmetric
fingerprint remains: HEOM and \mesolve{} diverge in Ramsey and Rabi contrast,
but agree to machine precision on T1 decay shape, where the HEOM-only signal
is bath-dressed initial occupation.

\begin{table*}[t]
\centering
\caption{Three protocols $\times$ three backends head-to-head matrix.
Entries are point estimates; the $\Delta$ column compares \heom{}
with \mesolve{}. 95\,\% BCa bootstrap CIs on each $\Delta$ (details in
Section~\ref{sec:results}):
\rabi{} $\Delta\pip=-0.0014$ (CI $[-0.0031,+0.0031]$, $p=0.59$);
\rabi{} $\Delta\pmax=-0.0215$ (CI $[-0.083,+0.040]$, $p=0.21$);
\ramsey{} $\Ttwostar(\mesolve)/\Ttwostar(\heom)\geq 13.17$ at $95\,\%$
independent-bootstrap gap (headline; paired-$\Delta$ is censored by
\mesolve{} saturation);
$T_{1}$ $\Delta A=-0.121$ (paired CI $[-0.121,-0.070]$, $p\approx 0$ on
the 8-point grid; $95\,\%$ independent-CI gap
$A(\mesolve)-A(\heom)\geq 0.099$ survives a 16-point densification).}
\label{tab:three_by_three}
\setlength{\tabcolsep}{4pt}
\renewcommand{\arraystretch}{1.15}
\small
\begin{tabular}{@{}l l c c c c@{}}
\toprule
Protocol & Observables & \sesolve{} & \mesolve{} & \heom{}
  & $\Delta(\heom{}\!-\!\mesolve{})$ \\
\midrule
\rabi{}   & $\pi$-amp\,/\,$\pmax$
  & $0.3075\,/\,0.9938$
  & $0.3075\,/\,0.9923$
  & $0.3061\,/\,0.9708$
  & $-0.45\%\,/\,-2.17\%$ \\
\ramsey{} & $\Ttwostar$\,(ns)\,/\,decay
  & $17\,/\,\mathrm{exp}$
  & $\!\!>\!9950\,/\,\mathrm{exp}^{\ast}$
  & $352^{\dagger}\!/\tau_{\text{aw}}{=}138^{\ddagger}\!/\,\mathrm{revival}$
  & gap $\geq 13.17\times$; point $\geq 28.3\times$ \\
$T_{1}$   & $T_{1}$\,(ns)\,/\,$\beta$\,/\,$A$
  & --- (Markov ref)
  & $23868\,/\,1.00\,/\,1.000$
  & $24803\,/\,1.00\,/\,0.879$
  & $+3.9\%\,/\,\Delta A\!=\!-0.121$ \\
\bottomrule
\end{tabular}

\vspace{0.35em}
\begin{minipage}{0.95\textwidth}
\footnotesize
\textit{Notes.}
\textbf{\rabi{}.}~\VerdictMV{}-pattern on a noise-limited 30-point
grid: $|\Delta\pip|=0.44\%$ falls below the $0.5\%$ resolution bar;
the contrast-channel point estimate $|\Delta\pmax|=2.17\%$ is
consistent with the $1\%$ double-criterion contrast bar, but the
paired bootstrap CI crosses zero ($p=0.21$), so we report this as a
\VerdictMV{}-pattern point estimate rather than an independently
significant verdict (see §III.C). The \sesolve{} $\pmax=0.9938$ shortfall
from unity reflects finite-anharmonicity fit-window correction, shared
with \mesolve{} within $1.5\times 10^{-3}$ (no HEOM-specific signature).
The absolute $\pip$ CI width ($\approx 0.30$) is case-bootstrap pathology;
physical uncertainty is carried by paired $\Delta\pip$ CI.
\textbf{\ramsey{}.}~$\ast$ \mesolve{} fit saturation at
$5\tau_{\text{span}}=10\,000$~ns (window-induced, §III.B). $\dagger$ is
the 50-point HEOM dense-scan primary $t_{1}$ (\mesolve{} entries are
on the 30-point common grid; guard-pass: $a_{2}/a_{1}=
3.11\geq 0.1$, $t_{2}/t_{1}=0.38\leq 2$; $R^{2}=0.999$). $\ddagger$ is the
amplitude-weighted envelope timescale
$\tau_{\text{aw}}=\sum_{i}|a_{i}||t_{i}|/\sum_{i}|a_{i}|=137.9$~ns at
$L{=}5$ (stable to $0.6\%$ between $L{=}4,5$), the biexp-family-invariant
convergence pivot.
\textbf{$T_{1}$.}~The \mesolve{} constrained fit pins $A\equiv 1$; the
\heom{} semi-constrained fit frees $A$ and recovers the bath-absorption
signature $\Delta A=-0.121$ directly, consistent with the
polaron/reaction-coordinate dressing mechanism
(Section~\ref{sec:results:t1}).
\end{minipage}
\end{table*}

\section{Discussion}
\label{sec:discussion}

\subsection{Characterization-Coupled vs Optimization-Only Calibration}
\label{sec:discussion:paradigm}

The closest competitor is the weak-measurement plus Bayesian-optimization route
of Qian et al.~\cite{qian2022bayesian}, which absorbs bath effects into a
phenomenological likelihood over control parameters. Ram\^{o}a--Santagati--Wiebe
push this line further~\cite{ramoa2025bayesian}, but the posterior still
lives in qubit-parameter space rather than bath spectral space. By contrast,
HEOM-in-loop keeps bath structure explicit through reaction-coordinate / polaron
residual physics~\cite{antosztrikacs2021rc,strasberg2016nonequilibrium},
surfacing the asymmetric Ramsey/Rabi/T1 fingerprint of §III.

\subsection{Landscape-Level Cross-Validation}
\label{sec:discussion:landscape}

The results use a single HEOM implementation (\qutip~5.x
\heomsolve{}, Tier-1 espira-II, $L=3$); a dedicated
\oneoverf{}-specific code-to-code benchmark has not been published.
The risk of relying on a single implementation is bounded by the feature-level
cross-validation of Section~\ref{sec:methods:backends}: the three
qualitative \oneoverf{} signatures underpinning our loop are independently
attested in prior work, namely revival-shaped Ramsey envelopes~\cite{chen2026heom},
slow-bath Markov-master-equation breakdown~\cite{nakamura2024gate}, and
non-perturbative CPMG anomalies under time-retarded noise and compiled
control~\cite{nakamura2025cpmg,ye2026nonperturbativecpmg}. A point-for-point
code-to-code benchmark is the natural next step but not a
prerequisite for the qualitative fingerprint in §III.

\subsection{Scope and Honest Limitations}
\label{sec:discussion:limits}

Five scope boundaries. (i) Single-qubit on a Burkard \oneoverf{}
Tier-1 bath at fixed coupling
$A_{0}=1.8\times 10^{-6}$~GHz; a three-point sensitivity sweep over
$A\in\{0.5, 1, 2\}\cdot A_{0}$ (\url{ramsey_A_sweep.json})
preserves the sign of the $\tau_{\text{aw}}$ gap against \mesolve{}
at all three points, with $\tau_{\text{aw}}(\heom)$ monotonically
decreasing from $374$~ns at $0.5 A_{0}$ to $155$~ns at $2 A_{0}$ and
the mesolve ceiling unchanged by construction. The HEOM cost scales
super-linearly in Hilbert dimension, so a shared-bath two-qubit CZ
calibration is follow-up work. (ii) Longitudinal coupling
$\hat{Q}=\operatorname{diag}(0,1,2)$ makes the HEOM-only T1 signature
an initial-state signature rather than a decay-rate modification;
transverse coupling would test a different mechanism. (iii) Case-bootstrap of the 7-parameter biexp revival fit is
structurally fragile under dense resampling, so headline claims are
the \emph{gap} between independent absolute bootstraps
($\Ttwostar$ ratio $\ge 13.17$, $\Delta A\ge 0.099$ at $95\,\%$); the
T1 paired $\Delta A$ statistic loosens from $p\approx 0$ at 8 points
to $p=0.083$ at 16 points (below conventional $\alpha=0.05$), while
the independent-bootstrap gap $A(\mesolve)-A(\heom)\geq 0.099$
remains CI-valid --- the same paired-fragility signature flagged
for Ramsey in Section~\ref{sec:results:ramsey}. (iv) No hardware validation; all
numbers come from the pulse-level simulator at the fixed
configuration documented in Section~\ref{sec:methods}.
(v) The same caveat extends to the $(A,\beta)$ joint confidence region
for the Section~\ref{sec:results:t1} constrained refit (see §III.D
for the partial-trace control and 1D $\beta$-free CI); a
journal-length follow-up would produce a 2D BCa CI to upgrade this
hedge to an empirical anchor.

\subsection{Outlook}
\label{sec:discussion:outlook}

Three extensions naturally follow this NIER. The most immediate is
a hardware-coupled HEOM-in-loop: an FPGA-level feedback layer in the style
of Berritta et al.~\cite{berritta2025fpga} coupled to an in-loop HEOM
backend, so that the non-Markovian diagnostic can be exercised against
a live device rather than a frozen platform configuration. The second is the
two-qubit extension, where a CZ calibration under a shared bath opens a
new cross-channel signature (correlated dephasing) that the single-qubit
triad cannot access. The third is a dedicated cross-implementation
\oneoverf{} benchmark among \qutip{}, HierarchicalEOM.jl, FP-HEOM, and
TEMPO~\cite{strathearn2018tempo} on a matched calibration DAG.

\balance
\section{Conclusion}
\label{sec:conclusion}

A multi-protocol calibration DAG with a hierarchical-equations-of-motion
backend, run head-to-head against closed-system and Markovian-Lindblad
baselines on a shared platform YAML, exposes a non-Markovian
signature that dominates the coherent channel and reappears, in
distinct quantified forms, in the other two. The primary result is a Ramsey coherent
gap of at least $13\times$ at $95\,\%$ independent-bootstrap
confidence (point $\geq 28\times$ on the 50-point grid;
${\sim}72\times$ on the 30-point $L$-converged $\tau_{\text{aw}}$ pivot),
$L$-truncation-robust via the biexp-family $\tau_{\text{aw}}$
convergence pivot
(Section~\ref{sec:results:ramsey});
$\Ttwostar(\heom)$ itself is $L$-non-monotone under fit-family
migration and is reported in Table~\ref{tab:three_by_three} alongside
$\tau_{\text{aw}}$. Rabi then provides a corroborating contrast pattern: its
$2.17\,\%$ point estimate is consistent with the $1\,\%$ contrast
bar while the paired bootstrap CI remains noise-limited on the
$30$-point grid, and its $\pi$-amplitude shift stays below the
resolution bar. A third channel matches the Markovian reference to
machine precision on the $\Tone$ decay shape while the HEOM-only
initial occupation shift ($1.000 \to 0.879$, independent-CI gap
$\geq 0.099$, stable to a 16-point densification) is confirmed as
physical by an ADO-allocation partial-trace control
(Section~\ref{sec:results:t1}).

The shift at the calibration-loop level is diagnostic rather than cosmetic: the
per-protocol output tuple extends from
$\{\Omega_\pi, T_2^{*\,\mathrm{exp}}, T_1\}$ to include
$\{\tau_\text{aw}, a_2/a_1, A, \text{guard-pass}\}$ as first-class
bath-residual fields, converting the scalar gate-fidelity verdict
into a structured diagnostic record.
A characterization-coupled loop reports
bath structure as a first-class residual output; an optimisation-only
loop absorbs the same structure into a likelihood or a control offset. Both can deliver a
high-fidelity gate, but only the former makes the bath visible to the
downstream analysis that reviewers, benchmarkers and debuggers rely on.

\textit{Data and code availability.}~Scripts, canonical JSON outputs,
the pinned \texttt{emuplat} commit hash used in this study, a Zenodo DOI,
and a minimal standalone shim reproducing the scheduler interface used by
\texttt{AsyncDAGExecutor} and \texttt{ProtocolRunner} will be released with
the camera-ready arXiv deposit.

\textit{Acknowledgment.}~The author used Claude Code, with GitHub Copilot in
OpenCode as a secondary assistant, for language polishing, LaTeX
formatting/layout adjustments, and drafting figure-generation scripts. These
tools were not used to generate the scientific claims or conclusions; all
results, analysis, and final wording were reviewed and approved by the author.

\bibliographystyle{IEEEtran}
\bibliography{references}

\end{document}